\documentclass[11pt]{article}
\usepackage[margin=1in]{geometry}
\usepackage[T1]{fontenc}
\usepackage{lmodern}

\usepackage[style=science, doi=true, backend=biber]{biblatex}
\usepackage{graphicx}
\usepackage{float}
\PassOptionsToPackage{hyphens}{url}
\usepackage{hyperref}
\hypersetup{colorlinks=false}
\usepackage{bookmark}
\usepackage[all]{hypcap}
\usepackage[version=4]{mhchem}
\usepackage[capitalise]{cleveref}

\usepackage{wrapfig}
\usepackage{sidecap}
\usepackage{subcaption}
\usepackage[labelfont={bf,footnotesize}, textfont=footnotesize, skip=8pt plus 2pt minus 2pt]{caption}

\usepackage{mathtools} % should load amsmath
\usepackage{amsmath}
\usepackage{amssymb}
\usepackage{amsfonts}
\usepackage{xspace}
\usepackage{changepage}
\usepackage{breqn}

\usepackage{cleveref}
\usepackage{textcomp}

\usepackage{lineno}
\usepackage{array}
\usepackage{booktabs}  % Nice tables!
\usepackage{multicol}
\usepackage{multirow}

\usepackage{adjustbox}

\usepackage{siunitx}
\DeclareSIUnit{\molecule}{molecule}
\DeclareSIUnit{\molecules}{molecule}
\DeclareSIUnit\molar{\mole\per\cubic\deci\metre}
\DeclareSIUnit\Molar{\textsc{M}}
\usepackage[shortlabels]{enumitem}
\setlist{nosep} 

\usepackage{xcolor}

\usepackage[color=green!20]{todonotes}

\newcommand{\thinsim}{{\raise.17ex\hbox{\(\scriptstyle\mathtt{\sim}\)}}}

\usepackage[flushleft]{threeparttable}

\begin{document}
\begin{centering}
\textbf{\Large Can biophysical models give insight into the synaptic changes associated with addiction? }\\[3mm]
\textbf{Mayte Bonilla-Quintana and Padmini Rangamani*}\\[1mm]
Department of Mechanical and Aerospace Engineering,\\
    University of California San Diego, La Jolla CA 92093.\\
    $^{*}$To whom correspondence must be addressed: prangamani@ucsd.edu\\

\end{centering}

\begin{abstract}

Effective treatments that prevent or reduce drug relapse vulnerability should be developed to relieve the high burden of drug addiction to society.
This will only be possible by enhancing the understanding of the molecular mechanisms underlying the neurobiology of addiction.
Recent experimental data have shown that dendritic spines, small protrusions from the dendrites that receive input from excitatory neurons, from spiny neurons in the nucleus accumbens exhibit morphological changes during drug exposure and withdrawal.
Moreover, these changes relate to the characteristic drug-seeking behavior of addiction.
However, due to the complexity of the dendritic spines, we do not yet fully understand the processes underlying their structural changes in response to different inputs.
We propose that biophysical models can enhance the current understanding of these processes by incorporating different, and sometimes, discrepant experimental data to identify the shared underlying mechanisms and generate experimentally testable hypotheses.
This review aims to give an up-to-date report on biophysical models of dendritic spines, focusing on those models that describe their shape changes, which are well-known to relate to learning and memory. 
Moreover, it examines how these models can enhance our understanding of the effect of the drugs and the synaptic changes during disease progression.

%Neurons are connected via synapses. 
%In some regions of the brain, the receiving part of the synapse, the postsynapse, forms a small protrusion from the dendrite, namely the dendritic spine.  
%In medium spiny neurons (MSNs) in the nucleus accumbens (NAc), these spines exhibit morphological changes during drug exposure and withdrawal.
%Since these changes relate to the characteristic drug-seeking behavior of addiction, it is important to understand the underlying neurobiological mechanisms. 
%However, this has not been an easy task because different drugs affect spines differently. 
%But more importantly, we do not yet fully understand the processes underlying the structural changes of dendritic spines in response to different inputs.
%We propose that biophysical models can enhance the current understanding of these processes by incorporating different, and sometimes, discrepant experimental data to identify the shared underlying mechanisms and generate experimentally testable hypotheses.
%This review aims to give an up-to-date report on biophysical models of dendritic spines, focusing on those models that describe their shape changes, which are well-known to relate to learning and memory.
\end{abstract}

%stronger connection between drugs and spine biophysics. This can be addressed by the following sections.
%1)subsection on how different drugs can impact the well-known signaling pathways
%2)diagrams of signaling pathways in healthy and addicted states
%3)experimental measurements that lend themselves to model constraints, validation and ways to test predictions.

\section{Introduction}

The highest burden of drug addiction to all societies is the immense loss of human capital -- from unfulfilled personal ambitions to loss of family structure. 
According to the National Institute on Drug Abuse \cite{abuse_overdose_2021}, in 2019 more than 70,000 Americas died from a drug-involved overdose, including illicit drugs and prescription opioids. 
However, this number only represents a small percentage of drug-related deaths.
According to a statistical study \cite{glei_estimating_2020}, in 2016 the drug-associated deaths was 2.2 times the number of drug-coded deaths.
Besides these losses, the estimated medical cost in hospitals of substance use disorder in 2017 was \$13.7 billion \cite{peterson_assessment_2021}.
%As of 2021, the economic cost of drug addition is 

A continued medical aspiration for relieving the high burden of drug addiction in society is to develop and tailor effective treatment to prevent relapse.
Assuming that drug addiction is a neuropsychiatric disease whose behavioral pathology consists in the propensity to relapse, even after periods of abstinence, then an effective treatment should prevent or reduce relapse vulnerability \cite{scofield_nucleus_2016}.
Because current behavioral and pharmacological therapy only helps a small percentage of patients \cite{scofield_nucleus_2016}, we need to understand the molecular mechanisms underlying the neurobiology of addiction. 
In the past decade, studies have focused on the neurons in the nucleus accumbens (NAc), a brain region that mediates goal-directed behavior, and their structural changes after drug exposure \cite{scofield_nucleus_2016, russo_addicted_2010, spiga_addicted_2014, gipson_relapse_2013}. 
These studies have noted that structural changes in neurons could be an important readout of addictive behavior.
For example, when animals are exposed to an environment associated with previous cocaine consumption, it triggers cocaine-seeking behavior that is induced by structural changes in NAc neurons \cite{gipson_relapse_2013}.
Another study showed that structural alterations to these neurons are only present in animals that show a persistent increase in the psychomotor activating effects of cocaine after repeated exposure to it, i.e., psychomotor sensitization \cite{li_induction_2004}.
However, there is no consensus on whether and how the structural changes in neurons derived from drug exposure drive addictive behavior   \cite{russo_addicted_2010,scofield_nucleus_2016} due to
opposing experimental data which show that the structural changes in NAc neurons after cocaine exposure are not required for locomotor sensitization and suggest that, instead, these changes may represent a compensatory process \cite{pulipparacharuvil_cocaine_2008}.
Complicating matters, studies have also shown that different drugs have different effects on these neurons \cite{scofield_nucleus_2016,russo_addicted_2010}.
For example, stimulants, such as cocaine, amphetamine, and nicotine, enhance the structural complexity of the neurons, while narcotics, like morphine, decrease it \cite{robinson_structural_2004}.

In this review, we explore the idea that experimentally-constrained biophysical computational models that relate structural changes of neurons to different drug conditions may be able to narrow down the possible mechanisms that may underlie the neurobiology of drug addiction. 
Although there are a repertoire of theoretical models of drug use and addiction, they rely on theories of how the brain solves computational, information-processing, and control problems \cite{gutkin_computational_2011,mollick_computational_nodate}.
Some of these models link these computations with specific brain areas \cite{mollick_computational_nodate}.
However, to our knowledge, they do not consider the structural and molecular changes of the neurons and, instead, focus on drug use behavior. 
Recently, biophysical models have informed the structural and molecular mechanisms related to learning and memory \cite{fauth_formation_2015,bonilla-quintana_exploring_2021,becker_biophysical_2021,miermans_biophysical_2017,bell_dendritic_2019,leung_systems_2021,ohadi_geometric_2019}. 
Here, we review such models and describe how they can broaden our understanding of the mechanisms behind addiction and relapse to aid the treatments for drug addicts. 
We focus on dendritic spines, small protrusions from the dendrites that receive excitatory input, and hence, form the postsynaptic part of the synapse.
Dendritic spines are thought to be standalone biochemical computational units \cite{svoboda_direct_1996,yuste_dendritic_1995,yuste_morphological_2001} and it is well-documented that they undergo structural plasticity during learning and memory formation \cite{lamprecht_structural_2004} and during drug consumption and withdrawal \cite{wright_silent_2020,gipson_relapse_2013,russo_addicted_2010}. 
Therefore, biophysical analyses of spines may provide us with the key links between these fundamental neurological processes.

This work begins by giving an overview of the structure and dynamics of dendritic spines. 
Then, we introduce the most studied forms of synaptic changes, namely, long-term potentiation (LTP) and long-term depression (LTD), and relate them with the modifications associated with drug addiction. 
After this biological background, we review different types of biophysical models and discuss how they can enhance our understanding of drug addiction and other neurological diseases. 
We finish by describing how these biophysical models can be extended to study memory formation and storage, and identifying the opportunities for predictive model development.

\section{Dendritic Spines}\label{sec:biology}
%\MBQ{The review will begin by giving an overview of the dendritic spines: their different classifications and the main structural components, namely, the actin cytoskeleton and the key receptors, which are involved in the synaptic changes related to addiction (Section \ref{sec:biology}).}

\subsection{Morphology}

Although dendritic spines were first described by Santiago Ramón y Cajal over a century ago, it has only recently become possible, through the development of novel imaging techniques, to investigate their structure-function relationships.
The dendritic spine is a bulbous protrusion from the dendritic shaft often connected by a thin neck, creating a spine head.
The postsynaptic density (PSD) is located at the tip of the spine head, close to the presynaptic terminal, and it is enriched with receptors, proteins, and signaling molecules (Fig. \ref{fig:bioscheme}A). 
The quantitative characteristics of the spine morphology are broadly described by: the length and width of its neck; the volume of its head; and the surface area of the PSD.
These geometric features alone have been studied in terms of correlations between the morphology and function of the spine \cite{arellano_ultrastructure_2007,ofer_ultrastructural_2021,tonnesen_spine_2014}.
For example, the spine head volume correlates with the area of the PSD, the number of postsynaptic receptors, the readily-releasable pool of transmitters, and the neck width \cite{arellano_ultrastructure_2007,ofer_ultrastructural_2021,tonnesen_spine_2014}.
Moreover, dendritic spines that connect to the same axon have similar size and head volume demonstrating that they share the same activation history \cite{bartol_computational_2015}.

Dendritic spines also have diverse shapes.
Thus, early studies classified them into three categories: 1) spines with long, thin necks and small heads (thin spines), 2) spines with thicker necks and large heads (mushroom spines), and 3) short spines without well-defined necks (stubby spines) \cite{peters_small_1970} (Fig.\ref{fig:bioscheme}B).
Experiments found that spine shapes were more distinct in adults than in young animals, which suggested that stubby spines form from the dendritic shaft when the axon contacts the membrane (shaft synapse) and then turn into thin or mushroom spines \cite{harris_three-dimensional_1992}.
However, the current view is that a filopodia-like structure that emerges from the dendritic shaft may connect to the axon to form a synapse \cite{zuo_development_2005,ziv_evidence_1996} evolving to a thin spine which later matures to a mushroom spine \cite{bourne_thin_2007}.
Moreover, thin spines are more prone to respond to synaptic activity and change their morphology accordingly, suggesting that they are learning spines that can mature to mushroom-like spines, which are more stable and thus, thought to be memory spines \cite{bourne_thin_2007}.
While mushroom spines can persist for months, thin spines only last a few days \cite{bourne_thin_2007}.
Dendritic spines have also been divided depending on the size of their heads \cite{kasai_structurestabilityfunction_2003}.
On one hand, spines with small heads (filopodia and thin spines) are motile and unstable. 
On the other hand, spines with large heads (mushroom and stubby spines) are stable and have stronger connections.
Thus, if we see spines as memory devices, large-headed spines are ``write protected'': they maintain preexisting connections and prevent new memory formation, while small-headed spines are ``write-enabled'' because they allow the acquisition of new memories \cite{kasai_structurestabilityfunction_2003}.

In addition to their unique morphological features, spines have attracted considerable attention of the signaling community because it has been demonstrated they compartmentalize calcium and that spine necks can filter membrane potentials, thereby isolating biochemical and electrical synaptic input from each other \cite{bloodgood_neuronal_2005}.
These findings have led to the hypothesis that spines implement input-specific learning rules \cite{arellano_ultrastructure_2007}. 
The shape changes in spines, in response to synaptic input, are possible by a reconfiguration of the spine cytoskeleton, mainly composed of actin filaments; actin remodeling is triggered by a cascade of chemical reactions due to calcium influx into the spine \cite{cingolani_actin_2008,honkura_subspine_2008,borovac_regulation_2018,konietzny_dendritic_2017,korobova_molecular_2010,hotulainen_actin_2010}. 
Changes in spine shape and size not only occur in development but also in adulthood  \cite{rochefort_dendritic_2012}, for example, during motor skill learning \cite{yu_spine_2011} and fear learning and extinction \cite{lai_opposite_2012}.
This dynamic change to spine morphology is known as structural plasticity \cite{nakahata_plasticity_2018}.
Note that due to this plasticity, the shapes of spines form a continuum instead of distinct categories \cite{ofer_ultrastructural_2021}.

\subsection{Structural Plasticity}\label{sec:plasticity}
%\MBQ{Secondly, the most studied forms of synaptic changes, long-term potentiation (LTP) and long-term depression (LTD), and the corresponding structural changes in the spine are introduced (Section \ref{sec:plasticity}).}

\begin{figure}[h]
    \centering
    \includegraphics[width=0.9\textwidth]{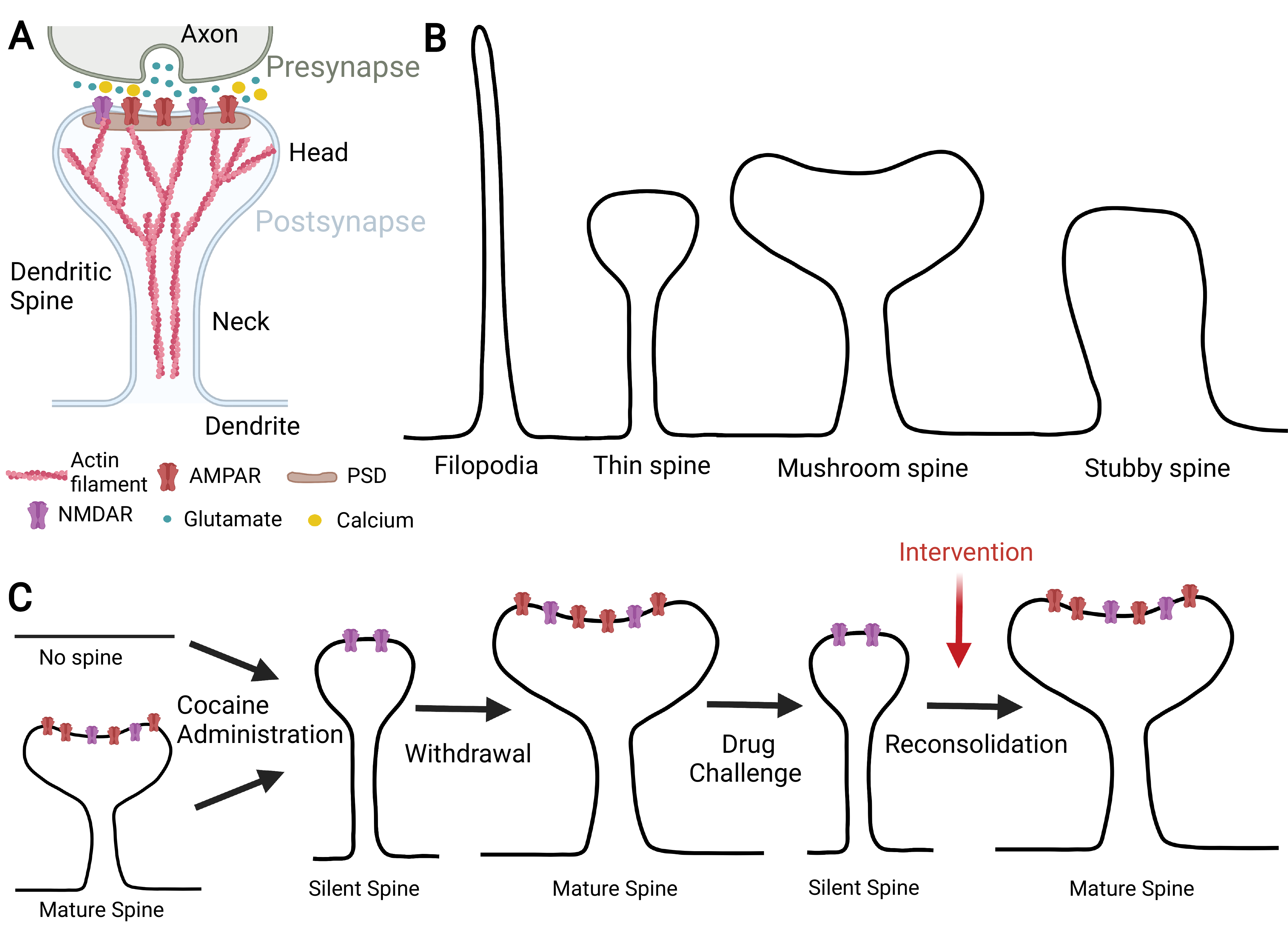}
    \caption{\textbf{Dendritic Spine}. A) Schematic depiction of a synapse. The dendritic spine (postsynapse) is in blue and the presynaptic bouton is in green. B) Different types of dendritic spines. C) Hypothesized evolution upon cocaine administration, withdrawal, and drug challenge, proposed by \cite{wright_silent_2020}. The red arrow signals the window when the synapse is destabilized and treatments can take place to avoid reconsolidation. 
    Created with \url{BioRender.com}}
    \label{fig:bioscheme}
\end{figure}

The connection between neurons can be strengthened or weakened in a process called synaptic plasticity, and this change in the connections has a direct impact on learning and memory formation \cite{murakoshi_postsynaptic_2012,rochefort_dendritic_2012,matsuzaki_structural_2004,zhou_shrinkage_2004}.
The most studied forms of these synaptic changes are long-term potentiation (LTP), which strengthens the synapse, and long-term depression (LTD), which weakens it.
In LTP, glutamate released due to the high-frequency stimulation of the presynaptic neuron is captured by the $\alpha$-amino-3-hydroxy-5-methyl-4-isoxazolepropionic acid receptors (AMPARs) located at the PSD of the dendritic spine (Fig. \ref{fig:bioscheme}A).
Due to this glutamate influx, the spine depolarizes and releases the Magnesium (Mg$^{2+}$) ion of the N-methyl-D-aspartate receptors (NMDARs), which allows the calcium (Ca$^{2+}$) influx into the spine and triggers a cascade of reactions that increase the number of AMPARs at the PSD and AMPARs mediated current.
Such increase facilitates the subsequent uptake of glutamate, and hence, the synapse strengthens.
Moreover, dendritic spines show an associated increase in size, thus, linking their function and morphology \cite{matsuzaki_structural_2004,yuste_morphological_2001,yuste_morphological_2001,nakahata_plasticity_2018,murakoshi_postsynaptic_2012}.
LTD also depends on the activation of NMDARs, but it is induced by low-frequency stimulation of the presynaptic cell, causing glutamate release.
However, the amount of glutamate is not enough to depolarize the postsynaptic cell and remove all the Mg$^{2+}$ ions blocking the NMDARs, but only a few.
Hence, the Ca$^{2+}$ influx is lower than in LTP and triggers a different set of chemical reactions resulting in a decrease in the number of AMPARs at the PSD, and spine shrinkage \cite{zhou_shrinkage_2004}. 

%How are LTP and LTD impacted by drugs?
How is plasticity impacted by drugs?
It has been shown that cocaine use impairs the induction of LTP in spines of medium spiny neurons (MSNs) in the NAc \cite{gipson_relapse_2013}.
These spines are depressed after exposure to cocaine, which generates silent synapses that lack AMPARs but have NMDARs \cite{scofield_nucleus_2016,gipson_relapse_2013,wright_silent_2020,russo_addicted_2010}.
The lack of AMPARs impairs LTP because these spines cannot respond to glutamate release.
Because silent synapses are highly abundant during development, it has been hypothesized that the brain returns to a more juvenile state after drug exposure, which may underlie pathological drug-seeking behaviors \cite{gipson_structural_2017}. 
Interestingly, AMPARs are reinserted into these silent synapses only when a stimulus related to cocaine consumption is present \cite{gipson_relapse_2013,wright_silent_2020}. 
This synaptic potentiation induces drug craving, and thus, may explain the high rates of relapse \cite{gipson_relapse_2013}.
Moreover, re-silencing these synapses via optogenetic removal of AMPARs inhibits relapse-like behaviors \cite{gipson_structural_2017}.
A recent study shows that there is a time window in which previously silent synapses that are potentiated due to exposure to cues related to drug consumption can be temporally destabilized and vulnerable to disruptions before being consolidated (becoming mature) \cite{wright_silent_2020}. 
Moreover, this window relates to the dynamics of signaling molecules that alter the spine cytoskeleton.
Preventing spine maturation during this window decreases cue-induced cocaine seeking, which can be used for anti-relapse treatments \cite{wright_silent_2020} (Fig. \ref{fig:bioscheme}C).

However, the precise neuronal and molecular substrates encoding the dynamics of drug memories have not been fully identified \cite{wright_silent_2020,scofield_nucleus_2016}.
Moreover, until recently, new imaging techniques allowed for the study of the mechanisms underlying the structural and molecular changes of the synapses \cite{rochefort_dendritic_2012}, but they remain incompletely understood due to the complexity of the synapse and the diverse temporal and spatial scales of the experiments and underlying events.

\subsection{Signaling Pathways associated with structural plasticity}

Upon LTP induction, there is an increase of the number of AMPARs in the PSD due to the Ca$^{2+}$/ calmodulin-dependent protein kinase II (CaMKII) mediated phosphorylation of AMPAR auxiliary protein stargazin, which allows extrasynaptic diffusive  receptors to bind to PSD95, a PSD molecule \cite{lisman_mechanisms_2012,choquet_linking_2018,bats_interaction_2007}. 
The exocytosis of extrasynaptic AMPARs into the plasma membrane depends on the rat sarcoma virus-extracellular signal-regulated kinase (RAS–ERK) pathway, RAB-GTPase proteins, soluble  N-ethylmaleimide-sensitive-factor attachment protein receptors (SNARE proteins), syntaxin 4 and 13, and the motor protein myosin V \cite{lisman_mechanisms_2012,patterson_ampa_2010,lee_activation_2009,choquet_linking_2018}.
The RAS-ERK pathway is stimulated by CaMKII upon LTP induction \cite{zhu_ras_2002,lee_activation_2009}.
Thus, CaMKII has a dual function upon LTP: it immobilizes extracellular AMPARs in the PSD and replenishes the pool of extrasynaptic receptors via exocytosis.
Moreover, CaMKII is one of the most important molecules for LTP.
It is activated by calcium-calmodulin, which in turn is activated by Ca$^{2+}$ influx into the spine through NMDARs and it can act as a protein switch due to its autophosphorylation property. % at T286.
Such a state lasts longer than calcium elevation, thus CAMKII acts as a biochemical integrator of the multiple calcium pulses during LTP induction \cite{lisman_mechanisms_2012}.

\begin{figure}[h]
    \centering
    \includegraphics[width=0.95\textwidth]{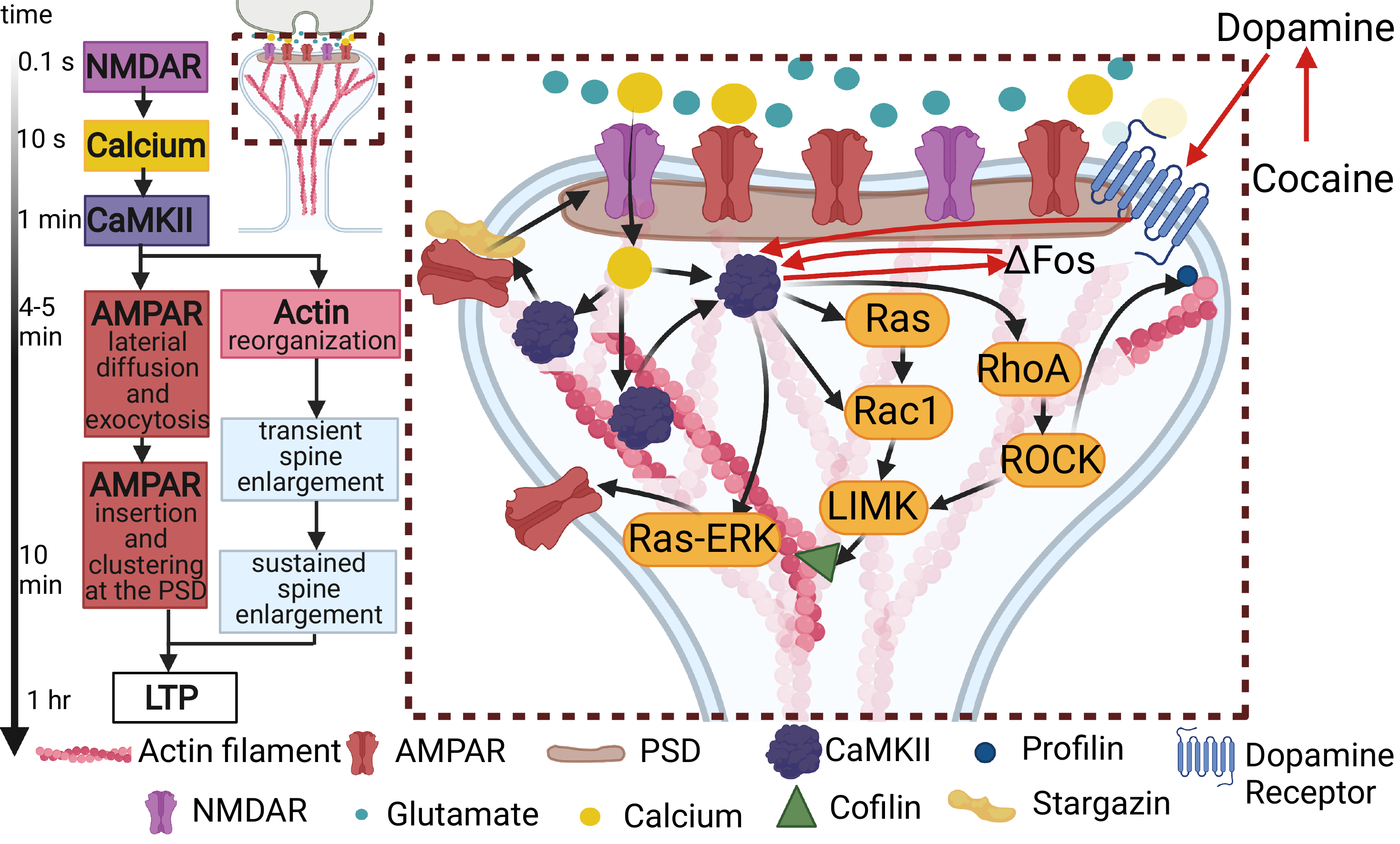}
    \caption{\textbf{Signaling pathways associated with LTP in the dendritic spine}.
    Right: events time-scale.
    Left: scheme of the signaling events in the dendritic spine.
    CaMKII is activated upon the Ca$^{2+}$ influx into the spine during LTP induction, which triggers the Ras, Rac1, and RhoA pathways. 
    CaMKII also induces AMPARs exocytosis via activation of the Ras-ERK pathway and capture of AMPARs into the PSD. 
    During cocaine consumption, dopamine receptors trigger CaMKII activity and $\Delta$Fos is increased. 
    Figure created with \url{BioRender.com}}
    \label{fig:signaling}
\end{figure}

In addition to its involvement with AMPARs, CaMKII also affects the dendritic spine cytoskeleton by stabilizing actin filaments \cite{okamoto_roles_2009}.
Actin is a globular protein (G-actin) that forms filaments (F-actin) and it is the main component of the spine cytoskeleton. 
F-actin is a polar structure that continuously polymerizes G-actin at its plus end and it is depolymerized at its minus end. 
In the spine, there is a dynamic equilibrium between G-actin and F-actin that is modulated by actin-binding proteins (ABPs) which promote F-actin depolymerization and G-actin polymerization \cite{okamoto_rapid_2004}.
Moreover, the actin cytoskeleton is restructured upon LTP by an orchestrated interplay between actin and ABPs \cite{bosch_structural_2014,okabe_regulation_2020}, which allows for spine enlargement.
In the basal state, F-actin is bundled by CaMKII, which stabilizes the bundle.
Upon NMDARs activation, CaMKII autophosphorylates and detaches from actin filaments, which allows CaMKII to interact with other molecules and remodel the actin cytoskeleton \cite{okamoto_roles_2009,wang_assemblies_2019}. 
Furthermore, CaMKII regulates actin dynamics by signaling pathways involving the Rho family of small GTPases, such as RhoA, Rac1, or Cdc42 \cite{okamoto_roles_2009}.
The activity of these GTPases is triggered by guanine-nucleotide-exchange factors (GEFs) and suppressed by GTPase activating proteins (GAPs), and they coordinately regulate ABPs activity.
For example, RhoA controls profilin II and cofilin \cite{maekawa_signaling_1999,witke_role_2004,okamoto_roles_2009}.
The former binds to G-actin and facilitates the polymerization of F-actin at its plus end. 
Cofilin, which is also controlled by Rac1, has a dual function depending on its concentration.
At low concentrations, cofilin severes F-actin and at high concentrations, it promotes F-actin nucleation \cite{andrianantoandro_mechanism_2006}.
Note that the nucleation and severing events produce more actin filaments, and the combination of both events is likely to be responsible for keeping a steady distribution of the F-actin length. 
Thus, cofilin regulation is critical for actin dynamics.
These pathways are depicted in Figure \ref{fig:signaling}.

It has been shown that cocaine decreases active RhoA in the NAc MSNs, which could lead to a decrease in cofilin \cite{kim_cocaine_2009}, affecting the spine cytoskeleton.
Moreover, activation of CaMKII by D1-like dopamine receptors in the NAc reinstates cocaine-seeking behavior; this is associated with an increase of AMPARs in the membrane \cite{anderson_camkii_2008}.
CaMKII also phosphorylates $\Delta$Fos and it is required for the cocaine-mediated accumulation of $\Delta$Fos in NAc \cite{robison_behavioral_2013} (red lines in Fig. \ref{fig:signaling}).
$\Delta$Fos is a Fos family transcription factor that shows long-lasting accumulation in the NAc after chronic administration of any drug of abuse, thereby supporting the view that changes in gene expression contribute to drug addiction. 
Interestingly, it has been shown that $\Delta$Fos upregulates the transcription of the CaMKII gene selectively in the NAc shell and that CaMKII stabilizes $\Delta$Fos, allowing for a greater accumulation of $\Delta$Fos that induces more CaMKII \cite{robison_behavioral_2013}.
However, how this positive feedback alters the spines is not fully understood, because of the complexity of the signaling pathways. 
Moreover, the signaling pathways activated by LTP induction have diverse temporal and spatial extents \cite{murakoshi_postsynaptic_2012}, and the coupling between the signaling pathways and spine structure is also complex (Fig.\ref{fig:signaling}). 
In order to disentangle the multiple spatial and temporal complexities involved in LTP induction and effects of drugs, we propose the theoretical models that incorporate biophysical details at various levels can shed light into the emergent behaviors of spines. 
These modeling approaches can elucidate the mechanisms underlying adaptation to addiction by allowing the incorporation of different experimental observations, eliminate possibilities that are not physicochemically feasible, and generate experimentally testable hypotheses.

\section{Biophysical Models of Dendritic Spines}\label{sec:models}

Addictive behaviors share common features with learning models \cite{jones_synaptic_2005}, and hence, learning theories \cite{magee_synaptic_2020} can be adapted to investigate these behaviors. 
However, in this work, we focus on the structural and functional changes in dendritic spines that can inform relapse propensity.  
We first present a review of the models that investigate signaling pathways in the dendritic spine relevant to the mechanisms involved in drug addiction, particularly those pathways that result in structural LTP or LTD due to an increase or decrease of AMPARs and CaMKII dynamics.
Then, we revise models that have incorporated idealized and realistic spine shapes linking the structural and functional characteristics of dendritic spines.  

\subsection{Signaling Models}\label{sec:signaling}
%\MBQ{Next, we review early signaling models of dendritic spines that describe their volume dynamics upon LTP (Section \ref{sec:signaling}). }

Early models of synaptic plasticity, based on Ca$^{+2}$ influx through the NMDARs, describe how different levels of Ca$^{+2}$ \cite{lisman_feasibility_1988,lisman_mechanism_1989} or induction protocols \cite{shouval_unified_2002} lead to an increase or decrease in the synaptic weight, which describes the strength of the synaptic connection. 
Later models provide a more detailed description of the biochemical reactions leading to either LTP or LTD \cite{pi_coupled_2008}, the volume regulation of dentric spines upon LTP \cite{rangamani_paradoxical_2016}, or the change in AMPARs mediated current promoting synaptic modifications  \cite{maki-marttunen_unified_2020}. 
These signaling models use ordinary differential equations (ODEs) to describe the biochemical reactions.
They span from simple descriptions based on key assumptions \cite{shouval_unified_2002} to detailed descriptions of the different chemical pathways \cite{maki-marttunen_unified_2020}.
As technological advances allow for the study of specific signaling pathways, models have incorporated these findings to test synaptic plasticity hypotheses \cite{ohadi_computational_2019}.
The increase in computational power has also allowed the use of agent-based models \cite{ordyan_interactions_2020}, in which the molecules are agents that follow a set of rules, mimicking chemical reactions leading to LTP.
Some of these models include the CaMKII activity \cite{pi_coupled_2008,lisman_mechanism_1989,ordyan_interactions_2020,rodrigues_stochastic_2021} and GTPases \cite{rangamani_paradoxical_2016,maki-marttunen_unified_2020}.

These models test different hypotheses based on the available experimental findings and create theories about how long-term information can be stored in the brain. 
For example, early models test the idea that CaMKII provides a bistable switch \cite{lisman_feasibility_1988} where ``off'' CaMKII is turned ``on'' when there is a synaptic event and stays on for long periods due to the autophosphorylation property of the kinase. 
Furthermore, this hypothesized switch can stay on for longer times than protein turnover because the newly synthesized unphosphorylated CaMKII subunit could become phosphorylated \cite{miller_regulation_1986}.
The mathematical models show the feasibility of this hypothesis \cite{lisman_feasibility_1988}, and expand it to add a phosphatase switch involved in LTD, providing a tristable system to examine how a synapse can be bidirectionally modified \cite{pi_coupled_2008}.
Models have also been used to systematically investigate how the interaction between CaMKII and other molecules can control the sensitivity of synapses to calcium signals \cite{ordyan_interactions_2020,rangamani_paradoxical_2016}.
Moreover, the incorporation of actin remodeling revealed that it added robustness to the dendritic spine response upon LTP \cite{rangamani_paradoxical_2016}.
One common difficulty for these models is parameter fitting since not all the reaction rates or protein concentrations are known in the spines, due to experimental difficulties. 
Thus, informed assumptions and decisions are made.
One common assumption is that reaction rates are the same across cell types, and protein concentrations are cell-type dependent; and hence, protein concentrations are fitted in the models \cite{maki-marttunen_unified_2020}. 
Although parameters are fitted when unknown and varied to see their effects in the model dynamics, they have to be within physiological regimes \cite{maki-marttunen_unified_2020,rangamani_paradoxical_2016}. 

As experiments began to investigate the source of the AMPARs that incorporate into the PSD upon LTP, i.e., whether they are exocytosed or laterally diffused into the PSD \cite{park_ampa_2018,makino_ampa_2009,park_recycling_2004,choquet_linking_2018}, theoretical studies examined the implications of the different sources to plasticity \cite{becker_biophysical_2021,burlakov_synaptic_2012,adrian_probing_2017,choquet_role_2003,bell_crosstalk_2021}.
Moreover, models started to explore how synapses are maintained over time periods that are longer than the average lifetime of these receptors and other molecules \cite{hayer_molecular_2005,shomar_cooperative_2017,shouval_clusters_2005}, and aided the study of AMPARs clusters formation \cite{nair_super-resolution_2013,shouval_clusters_2005}.
%%%%
These models propose different hypotheses for memory maintenance.
Instead of relying on a bistable molecular switch, like that of CaMKII, a model of clusters of AMPARs show that they are metastable with longer lifetimes than that of a single receptor,  and that synaptic weights, which depend on the number of AMPARs, can be bidirectionally changed \cite{shouval_clusters_2005}. 
An alternative hypothesis is that AMPARs exhibit a self-sustained switch related to their movement to and from the membrane \cite{hayer_molecular_2005}.
In this model, the presence of receptors at the synapse enhances the recruitment of more AMPARs, resembling the conversion from silent to active synapses \cite{hayer_molecular_2005}.
Moreover, a model based on simplified biophysical assumptions shows cooperative binding and unbinding of proteins at the PSD, replicating experimental observations \cite{shomar_cooperative_2017}.
By performing model simulations, the interplay between AMPARs and CaMKII dynamics has been examined \cite{hayer_molecular_2005,bell_crosstalk_2021}.
Another important player of structural plasticity, the actin filaments dynamics \cite{cingolani_actin_2008,borovac_regulation_2018} and its dynamic equilibrium \cite{okamoto_rapid_2004}, has also been investigated providing a quantitative description of the change of the filamentous to globular actin ratio upon LTP observed experimentally \cite{bennett_model_2011}. 
Moreover, computer simulations have aid the understanding of the F-actin binding and unbinding to CaMKII \cite{wang_assemblies_2019}.

\subsection{Models Incorporating Spine Shape}\label{sec:simple}
%\MBQ{Then, we present dendritic spine models that use simplistic geometries to represent the spine shapes (Section \ref{sec:simple}).}
Given the structural and morphological complexity of spine shapes, the incorporation of geometrical detail is a critical aspect of computational modeling for LTP.
Models started examining the role of the spatial localization of signaling molecules in synaptic plasticity \cite{kim_colocalization_2011, oliveira_subcellular_2012,ohadi_geometric_2019}, the effect of their diffusion on nearby spines \cite{oliveira_subcellular_2012,li_induction_2016,leung_systems_2021} or the localization of other Ca$^{2+}$ sources, for example the endoplasmatic reticulum or the mitochondria \cite{leung_systems_2021,bell_dendritic_2019,cugno_geometric_2019,bartol_computational_2015}, or astrocytes, a glia type cell of the central nervous system that contacts the synapse \cite{denizot_stochastic_2021}.
Many of these models use idealized, simplified geometries where the spine head is represented by an ellipsoid or a sphere and the spine neck by a cylinder connecting the spine head with the dendrite \cite{bell_dendritic_2019,leung_systems_2021,ohadi_computational_2019,kim_colocalization_2011,oliveira_subcellular_2012}.
The dynamics of the signaling molecules are described by reaction-diffusion models that use partial differential equations (PDEs).
Because some molecules are scarce in the dendritic spines, they cannot be described by their concentration amount, as is done for deterministic reaction-diffusion models, and thus, stochastic simulations were implemented to better represent their dynamics \cite{kim_colocalization_2011,oliveira_subcellular_2012}. 
Also, the localization and diffusion of AMPARs were examined in idealized geometries \cite{kusters_shape-induced_2013,adrian_probing_2017}.
Recent models have used reconstructed dendritic segments with spines from electron micrographs to give an insight into how the shape of dendritic spines influences calcium dynamics \cite{bartol_computational_2015,lee_3d_2020,holst_stochastic_2021}. 

Overall, the incorporation of spatial detail in the models allowed examination of how molecular organization in the dendritic spines affect the response of signaling molecules to calcium \cite{ohadi_geometric_2019}, and the function of proteins \cite{kim_colocalization_2011}.
Moreover, different hypotheses, that are not possible to experimentally test yet, about how organelle size and contact between organelles affect Ca$^{2+}$ dynamics have been tested \cite{leung_systems_2021,bell_dendritic_2019}.
Although including the spatial component in the models increases their complexity and the computational burden of the simulations, it is necessary to shed light on mechanisms that rely on the protein distribution and translocation, as structural plasticity.

Models started to explore the role of actin cytoskeleton remodeling into spine maturation, how different configurations of actin filaments derive in the different types of spines due to a balance between a force generated by actin polymerization and the resistance offered by the spine membrane \cite{miermans_biophysical_2017,alimohamadi_mechanical_2021}, and show the importance of the spine-neck constriction by structures like actin rings \cite{bar_periodic_2016} for spine stability.
These models assumed isotropic forces generated by actin pushing the lipid membrane forward, which is represented as a thin elastic shell that minimizes its bending energy \cite{helfrich_elastic_1973}, to investigate the equilibrium configurations of symmetric spine shapes.

However, models that incorporate spatio-temporal dynamics of signaling often neglect the dynamics of actin remodeling because of the inherently different types of governing equations.
Signaling models usually focus on reaction and diffusion in a spatio-temporal manner \cite{bell_dendritic_2019,ohadi_geometric_2019,leung_systems_2021,kim_colocalization_2011,oliveira_subcellular_2012}.
Models incorporating actin remodeling usually involve force balances and movement of the boundary (spine membrane in this case) in response to those forces \cite{alimohamadi_mechanical_2021,miermans_biophysical_2017}.
A model coupling the signaling spatio-temporal dynamics and the membrane dynamics resulting from a force imbalance is hard to analyze because it is not always clear how its steady state is defined: the protein location affects the spine shape that, in turn, affects the protein dynamics.
Moreover, in basal conditions, proteins are not steady, they diffuse and react with other molecules, and the spine shape fluctuates \cite{fischer_rapid_1998,frost_single-molecule_2010}. 
This increases the complexity of computational simulations because, at each time step, the chemical reactions and spine shape have to be updated. 
Nonetheless, some efforts have been made to push the boundaries on these biophysical models forward. 
New models were proposed to investigate these spontaneous spine shape fluctuations \cite{bonilla-quintana_modeling_2020,bonilla-quintana_reproducing_2021} and their implications in LTP \cite{bonilla-quintana_exploring_2021,bonilla-quintana_reproducing_2021} by continuously changing the membrane according to the actin dynamics.
These models couple the biochemical and mechanical properties of dendritic spines, and thus, their function and structure.
This allows the comparison with time-lapse microscopy data and shed light into the underlying mechanisms for spine maintenance and changes upon LTP. 
Moreover, it promotes testing different hypothesis regarding to the most efficient mechanisms behind synaptic function.  

%\subsection{Models incorporating Real Dendritic Spine Shape}\label{sec:real}
%\MBQ{Finally, models are revised using asymmetric spine shapes and reconstructed spines from electron microscopy images (Section \ref{sec:real}). }

\subsection{Design considerations for model building}\label{sec:comparison}
%\MBQ{We compare these models and discuss the need to validate them with real spine shapes, and the trade-off between model detail level, insight, and computational burden. 
%Moreover, we highlight the advantages and disadvantages of each modeling scheme beyond the geometric representation of the spines, for example, deterministic versus stochastic description for the dynamics (Section \ref{sec:comparison}).}

Although we focus on models of dendritic spines, they span across different temporal scales. 
For example, Ca$^{+2}$ dynamics are fast (tens to hundreds of milliseconds) \cite{bartol_computational_2015,holst_stochastic_2021,cugno_geometric_2019,bell_dendritic_2019,shouval_unified_2002}
while kinases like CaMKII or more complex signaling dynamics are slower (seconds to minutes) \cite{pi_coupled_2008,maki-marttunen_unified_2020,ohadi_computational_2019,ohadi_geometric_2019,ordyan_interactions_2020,hayer_molecular_2005,kim_colocalization_2011,oliveira_subcellular_2012,leung_systems_2021}.
AMPARs dynamics also span from seconds, in studies of AMPARs lateral diffusion \cite{adrian_probing_2017}, to minutes, when investigating different sources of AMPARs \cite{becker_biophysical_2021,kusters_shape-induced_2013}.
Moreover, models investigating spontaneous fluctuations of spine shape \cite{bonilla-quintana_modeling_2020,bonilla-quintana_reproducing_2021} or AMPARs \cite{shomar_cooperative_2017} can span from seconds to minutes, respectively (Fig. \ref{fig:signaling}).

The level of detail also varies across studies, from simple models based on a few assumptions \cite{shouval_unified_2002} to detailed models accounting for all the signaling pathways \cite{maki-marttunen_unified_2020}. 
Most of the models look at the reaction induced by Ca$^{2+}$ influx through NMDARs, but also other sources of calcium such as organelles \cite{bell_dendritic_2019,leung_systems_2021}.
Moreover, the spatial representation of the dendritic spine also expands from a volume \cite{rangamani_paradoxical_2016} or number of AMPARs readout \cite{maki-marttunen_unified_2020} to the embodiment in static idealized \cite{bell_dendritic_2019,bartol_computational_2015,leung_systems_2021,adrian_probing_2017,miermans_biophysical_2017} and real geometries \cite{lee_3d_2020} or motile  shapes \cite{bonilla-quintana_exploring_2021,bonilla-quintana_modeling_2020,bonilla-quintana_reproducing_2021}.

The dynamics are described in different ways depending on the studied phenomena.
To select which type of model is the most suited to investigate a biological phenomenon, it is critical to know the readout of the model and whether it is comparable with experiments.
If there are no available experiments, at least there has to be evidence that such experiments can be performed with the current technology. 
This way, the model predictions can be tested, and the model can be used to motivate more experiments.
Most signaling pathways use ODEs to describe chemical reactions  \cite{ohadi_computational_2019} or PDEs when accounting for molecular diffusion \cite{ohadi_computational_2019}.
When the number of the studied molecules inside the spine is low, stochastic ODEs or PDEs are used instead \cite{kim_colocalization_2011,oliveira_subcellular_2012}. 
Other models take molecules \cite{ordyan_interactions_2020} or filaments \cite{bennett_model_2011} as agents that follow experimentally verified rules that mimic their reactions.

Overall, there is always a trade-off between the level of detail, the spatial and temporal scale, and computational burden when proposing theoretical models.
For example, for modeling the relapse propensity that lasts from hours to days \cite{wright_silent_2020}, most of the fast molecular dynamics can be assumed to reach a semi-steady state.
Hence, the model would only account for deviations of this state due to drug exposure.
In this top-down approach, only chemical pathways affected by drug consumption are considered, and the rest are assumed to be unaffected.
The affected pathways involve cytoskeleton proteins which induce remodeling. 
Thus, not only the variation of the AMPARs should be included, but the spine size and shape changes.
Moreover, mechanical properties, such as surface area tension that regulates cell functions and triggers signals \cite{gauthier_mechanical_2012}, should be considered in the models.

\section{Perspective: Can we use models to enhance our understanding of synaptic changes during disease progression and to study the effect of drugs?}\label{sec:drugs}
%\MBQ{We finalized the review explaining how these biophysical models incorporating realistic geometries can potentially be employed to enhance our understanding of different biological phenomena, especially the synaptic changes due to drug addiction to inform how to prevent relapse vulnerability (Section \ref{sec:drugs}).}

We consider that biophysical models can act as a computational microscope that incorporates diverse experimental findings to perform \textit{in silico} experiments and test hypotheses that are not yet accessible  experimentally. 
Therefore, they provide a platform to study the molecular effects of drug consumption in dendritic spines and simulate the impact of relapse treatments that inhibits or enhance certain signaling pathways. 
Although currently there is vast experimental measurements that can be use to restrict these models, these are sensitive to the the different experimental setups of drug administration: whether the drug is self-administrated or administrated by an experimenter \cite{scofield_nucleus_2016,russo_addicted_2010}.
Thus, the models should also consider these discrepancies. 
Models should be suited to incorporate different spatial and temporal extents of the signal transduction underlying LTP \cite{murakoshi_postsynaptic_2012} that is affected by cocaine.  
As a modeling community, this can be achieved by building modular, shareable models for different time- and spatial scales that can be coupled.
By developing biophysical models to understand the changes in the synapse during drug abuse, we can gain insight into mechanisms that may be involved in other neurological diseases. 
For example, dendritic spine alterations have been found in autism spectrum disorders, schizophrenia, and Alzheimer’s disease \cite{penzes_dendritic_2011,gipson_structural_2017}.
Such alterations correlate with the expression of proteins associated with the cytoskeleton \cite{bellot_structure_2014}, or the number of AMPARs \cite{zhang_bidirectional_2020}.

The models presented here can be extended to study the complex interplay between membrane mechanics, cytoskeleton dynamics, and protein interaction during endo- and exocytosis of extracellular vesicles (EVs) containing AMPARs that are inserted and removed from the spine membrane, respectively.  
Although such mechanisms have been modeled for other types of cells \cite{akamatsu_principles_2020,foret_kinetic_2008,noguchi_vesicle_2021}, they have not been incorporated into dendritic spine studies.
Furthermore, EVs containing proteins and RNAs are also endo- and exocytosed.
Recent evidence shows EVs are involved in pathologies of neurological diseases \cite{you_emerging_2019,holm_extracellular_2018}, and that they can potentially serve as diagnostic tools \cite{kano_extracellular_2019,you_emerging_2019,holm_extracellular_2018} or as vehicles for medicine delivery \cite{elsharkasy_extracellular_2020}.
Therefore, biophysical models can be used to complement the efforts in understanding the underlying mechanisms to propose new avenues to diagnose and treat neurological diseases.  

\section{Outlook}\label{sec:outlook}
% we have assume that a single spine is cappable to store memories, but also memories affect different brain regions 

We have assumed that dendritic spines are capable of storing memories: they protrude from dendrites and then connect to presynaptic sites. 
After LTP, the number of AMPARs at the PSD increase, enhancing the AMPAR-mediated conductivity and enlarging the spine. 
These mature spines can last over days or disappear to promote the creation of new memories, allowing for dynamic memory \cite{kasai_structural_2010}.
But the idea that memories reside on synapses has not been conclusively proven yet \cite{abraham_is_2019}.
Moreover, synaptic proteins have half-lives of a few days and there is a continuous turnover of spines \cite{abraham_is_2019,minerbi_long-term_2009}, but memories can last years. 
This leads to an alternative hypothesis stating that memories are rather stored in engrams, that are populations of neurons activated by an experience, across different brain regions \cite{josselyn_memory_2020}.
There have been modeling efforts to understand these different mechanisms, from the study of different learning rules that dictate the strength changes of the connection between neurons \cite{sjostrom_spike_2002,magee_synaptic_2020} to investigate how memories can be stored and recalled in different cell assemblies \cite{luboeinski_organization_2021} or how these networks can learn without disrupting memories \cite{tsuda_modeling_2020}.
Besides spanning across the spatial scale of the synapse and cell assemblies, the mechanisms underlying learning and memory storage span across different time scales \cite{tetzlaff_time_2012}.
Although models are addressing how memories can be stored despite high spine turnover \cite{fauth_formation_2015} or how multiple spines in the dendrite contribute to single neuron computations and how it is affected by dendritic morphology \cite{poirazi_illuminating_2020,poirazi_pyramidal_2003}, more studies are needed to address the pressing problems facing society.

\section*{Acknowledgements}
We thank members of the Rangamani Lab for discussion about dendritic spines.
This work was supported by an Air Force Office of Scientific Research Grant FA9550-18-1-0051 to P.R.

\printbibliography

\end{document}